\documentclass[12pt]{article}

\usepackage{putex}
\usepackage{autobreak}

\usepackage{tikz-feynman}
\usetikzlibrary{external}             
\immediate\write18{mkdir -p pgf-img}  
\usepackage[style=phys,biblabel=brackets,eprint=true]{biblatex}
\bibliography{mybib.bib}

\usepackage{graphicx}
\usepackage{caption}
\usepackage{amsmath}
\usepackage{array}
\usepackage{subcaption}
\usepackage{epstopdf}
\usepackage{enumerate}
\usepackage{youngtab}
\usepackage{tensor}
\usepackage{slashed}
\usepackage[aligntableaux=center]{ytableau}
\usepackage[utf8]{inputenc}
\usepackage{rotating}
\usepackage{multirow}
\usepackage[
      colorlinks=true,
      linkcolor=blue,
      urlcolor=blue,
      filecolor=black,
      citecolor=red,
      ]{hyperref}

\newcommand {\be} {\begin {equation}}
\newcommand {\ee} {\end {equation}}

\newcommand {\bes} {\begin {equation*}}
\newcommand {\ees} {\end {equation*}}

\newcommand{\cO}{{\mathcal O}}

\newcommand{\beq}{\begin{equation}}
\newcommand{\eeq}{\end{equation}}

\def\ie{\begin{equation}\begin{aligned}}
\def\fe{\end{aligned}\end{equation}}

\numberwithin{equation}{section}

\def\<{\langle}
\def\>{\rangle}

\newcommand{\ak}{\alpha}        
\newcommand{\bk}{\beta}         
        \newcommand{\Gk}{\Gamma}
\newcommand{\dk}{\delta}        \newcommand{\Dk}{\Delta}
\newcommand{\ek}{\varepsilon}   
\newcommand{\zk}{\zeta}

\newcommand{\lk}{\lambda}       \newcommand{\Lk}{\Lambda}
\newcommand{\rk}{\rho}          
\newcommand{\sk}{\sigma}        
          
\newcommand{\fk}{\phi}

\newcommand{\fcy}[1]{\mathcal{#1}}

\newcommand{\nb}{\partial}

\begin{document}

\institution{PU}{Joseph Henry Laboratories, Princeton University, Princeton, NJ 08544, USA}
\institution{FHI}{Future of Humanity Institute, University of Oxford, Trajan House, Mill Street, \cr  Oxford, OX2 0DJ, UK}

\title{
The Cubic Fixed Point at Large $N$
}

\authors{Damon J.~Binder\worksat{\PU,\FHI}}

\abstract{
By considering the renormalization group flow between $N$ coupled Ising models in the UV and the cubic fixed point in the IR, we study the large $N$ behavior of the cubic fixed points in three dimensions. We derive a diagrammatic expansion for the $1/N$ corrections to correlation functions. Leading large $N$ corrections to conformal dimensions at the cubic fixed point are then evaluated using numeric conformal bootstrap data for the 3d Ising model.}
\date{}

\maketitle

\tableofcontents

\section{Introduction}
\label{intro}

Theories often simplify when there are a large number of fields. For theories with vector-like large $N$ limits, the infinite $N$ behavior can often be solved exactly, and finite corrections computed in a systematic $1/N$ expansion. The most famous such class of theories are the $O(N)$ vector models, as reviewed for instance in \cite{Moshe:2003xn}, which have been studied in great detail using both traditional Feynman diagram methods and analytic bootstrap techniques \cite{Lang:1990ni,Petkou:1994ad,Maldacena_2013,Turiaci:2018nua}. Other theories which can be solved at large $N$ include scalar $O(M)\times O(N)$ and $S_M\ltimes O(N)^M$ models \cite{Kawamura:1988zz,Henriksson:2021lwn}, the Gross-Neveu-Yukawa model \cite{PhysRevD.10.3235} and its supersymmetric generalizations \cite{Gracey:1990aw}, conformal QED \cite{Appelquist:1988sr}, and the Abelian Higgs model \cite{LUSCHER1978465}. In each case, a renormalization group flow exists between a free field theory in the UV and an interacting fixed point in the IR. The flow becomes tractable at infinite $N$, and $1/N$ corrections to correlators in the IR can be computed systematically. Similar flows exist in holographic theories when the boundary is perturbed by a ``double-trace'' deformation, which changes  the boundary conditions of the bulk fields \cite{Witten:2001ua,Klebanov:2002ja}.

Techniques used to solve vector-like large $N$ limits have rarely been applied to cases where the UV theory is interacting. At strictly infinite $N$ it has long been known that the IR theory is almost identical to the UV theory, up to a shift in the conformal dimension of the perturbing operator \cite{PhysRev.176.257,Stanley:1968gx,PhysRevLett.31.1494}. Computing finite $N$ corrections to this limit, however, requires a quantitative understanding of the initial UV theory, and beyond $d=2$ this has proven challenging to acquire. Over the last decade, the numeric conformal bootstrap has emerged as a practical tool for non-perturbative computations in interacting conformal field theories \cite{Rattazzi:2008pe,Poland:2018epd}. For the Ising model in $d = 3$, the numeric bootstrap provides the most precise determinations of both conformal dimensions and OPE coefficients \cite{ElS12,El-Showk:2014dwa,Kos:2014bka,Kos16}:
\begin{equation}\begin{aligned}
\Dk_\sk &= 0.5181489(10)\,,\qquad &\Dk_\epsilon &= 1.412625(10)\,, \\
f_{\sk\sk\epsilon} &= 1.0518537(41) \,,\qquad &f_{\epsilon\epsilon\epsilon} &= 1.532435(19)\,, \\
\end{aligned}\end{equation}
where the error bars are rigorous bounds. Further, though less rigorous, results are available for many other conformal dimensions, OPE coefficients, and even some four-point functions \cite{Simmons-Duffin:2016wlq,Rychkov:2016mrc}. Computations competitive with the best Monte Carlo results for critical exponents have been recently achieved for the $O(2)$ and $O(3)$ models \cite{2015Kos,Kos16,Chester:2019ifh,Chester:2020iyt}, while also yielding OPE coefficients which are otherwise difficult to determine using traditional methods.

Our aim in this paper is to use the 3d Ising model bootstrap data to study the cubic fixed points at large $N$. We will do so by coupling $N$ Ising models via the interactions 
\begin{equation}\label{CubicCouple}
L_{\text{int}} = \int d^dx\,\left(-\mu\sum_i\epsilon_i(x) - \frac \lk2\sum_{i\neq j}\epsilon_i(x)\epsilon_j(x)\right)\,,
\end{equation}
where $\epsilon_i(x)$ is lowest dimension $\mathbb Z_2$-even operator in $i^{\text{th}}$ copy of the 3d Ising model. These are the only relevant interactions preserving the $S_N\ltimes \mathbb Z_2^N$ hypercubic symmetry of the $N$ decoupled Ising models. As we flow into the IR while suitably fine-tuning $\mu$, the theory flows to a conformal field theory known as the cubic fixed point \cite{Aharony:1973zz,PhysRevLett.31.1494}. Our strategy is similar to that in \cite{Komargodski:2016auf,Behan:2259464}, in which conformal perturbation theory was combined with bootstrap data in order to study the random-bond Ising model (which is the $N\rightarrow0$ limit of the cubic fixed point) and the long-ranged Ising model respectively. Unlike conformal perturbation theory, in which a controlled approximation is possible only given an exactly marginal operator, the large $N$ limit naturally gives us a controlled approximation for the cubic fixed points without this onerous requirement.

Recall that the cubic model is described within the $4-\epsilon$ expansion by the Lagrangian
\begin{equation}\label{Lag}
L = \int d^dx\,\left(\frac12\sum_{i=1}^N (\nb\fk_i)^2 + m^2\fk_i^2 +\frac{g}4\left(\sum_{i=1}^N\fk_i^2\right)^2 +  \frac{h}{4!}\sum_{i=1}^N \fk_i^4\right)\,,
\end{equation}
which is the most general renormalizable Lagrangian invariant under $S_N\ltimes \mathbb Z_2^N$ hypercubic symmetry. To leading order in the $4-\epsilon$ expansion there are four fixed points \cite{Aharony:1973zz}:
\begin{enumerate}
\item The free field fixed point, with $g = h = 0$.
\item $N$ decoupled Ising models, with $h\neq0$ and $g=0$.
\item The $O(N)$ fixed point, with $g\neq0$ and $h=0$.
\item The cubic fixed point, with $g\neq0$ and $h\neq0$.
\end{enumerate}
The first two fixed points are never stable, while the cubic fixed point is stable for all $N>N_c(\epsilon)$, and the $O(N)$ model stable otherwise. The precise value of $N_c$ depends on the spacetime dimension, but for $d=3$ it has been recently shown to be less than $3$ \cite{Chester:2020iyt} using the numerical conformal bootstrap, following previous attempts to estimate it using perturbative expansions \cite{PhysRevB.56.14428,Carmona:1999rm,Adzhemyan:2019gvv}. The $O(3)$ fixed point is thus unstable to cubic anisotropies. Such anisotropies are generically present in real crystals, and so second-order phase transitions in these materials are expected to be described by the cubic fixed point rather than the $O(3)$ fixed point.

Our plan is as follows. In Section~\ref{GENFUNC} we introduce a Hubbard-Stratonovich field $\fk(x)$ to simplify the interaction \eqref{CubicCouple}, allowing us to derive a path-integral for the generating functional of the cubic fixed point. At large $N$ the path-integral for $\fk(x)$ is dominated by a non-local Gaussian kinetic term, along with an infinite series of non-local higher-point interactions suppressed by increasing powers of $1/N$. Correlators of $\fk(x)$ can thus be computed using a Feynman diagram expansion, and with a little more work these diagrams can be extended to include other local operators.

Having derived a systematic expansion for large $N$ correlators at the cubic fixed point, in Section~\ref{ANOM} we compute the leading corrections to the conformal dimensions of various local operators at the cubic fixed point. Some of these calculations can be performed exactly, while others require integrating certain four-point functions in the 3d Ising model numerically. We close with a discussion of our results in Section~\ref{DISC}, and an appendix then follows.

\section{Generating Functional for the Cubic Fixed Point}
\label{GENFUNC}

Our ultimate goal is to compute correlators at the cubic fixed point. To do so, let us introduce source fields $J_i(x)$ and $K_i(x)$ coupled to $\epsilon_i(x)$ and $\cO_i(x)$ respectively, where $\cO_i(x)$ is some other local operator in the 3d Ising model. We are most interested in computing correlators of $\sk_i(x)$, where $\sk(x)$ is the lowest dimension parity odd operator in the Ising model, but our method is general and allows us to compute correlators of any local operator. 

Let us begin by defining the generating functional for correlators of $\epsilon_i(x)$ and $\cO_i(x)$:
\begin{equation}\label{initGen}\begin{split}
Z_{\lk,\mu}&[J_i,K_i] = \left\<\exp\left(\sum_{i=1}^N J_i(x)\epsilon_i(x) + K_i(x)\sigma_i(x)\right)\right\>\\
&= \left\<\exp\left(\int d^dx\,\left[-\frac \lk2\sum_{i\neq j}\epsilon_i(x)\epsilon_j(x)+\sum_{i=1}^N J_i(x)\epsilon_i(x) + K_i(x)\cO_i(x)\right]\right)\right\>_0\,,
\end{split}\end{equation}
where we use $\<\cdot\>_0$ to denote expectation values in the decoupled theory. The Hubbard-Stratonovich trick allows us to rewriting the cubic coupling using an auxiliary field:
\begin{equation}\begin{split}\label{HSTrick}
\exp&\left(\int d^dx\,\left[-\frac \lk2\sum_{i\neq j}\epsilon_i(x)\epsilon_j(x)\right]\right)\\
&\qquad \propto \int D\phi(x) \exp\left(\int d^dx\,\left[\frac1{2\lk}\fk(x)^2 + \fk(x)\sum_{i=1}^N\epsilon_i(x)\right]\right)\,.
\end{split}\end{equation} 
Note that to properly make sense of this equation we must regulate the $\fk(x)$ kinetic term, a subtlety we discuss in Appendix \ref{REGHS}. Applying \eqref{HSTrick} to our generating functional and then redefining ${\fk(x)\rightarrow\fk(x) + \mu}$, we find that 
\begin{equation}\label{gen2}\begin{split}
Z_{\lk,\mu}[J_i,K_i] &\propto \int D\fk \exp\left(\frac1{\lk}\int d^dx\,\left[\frac1{2}\fk(x)^2 + \mu\fk(x)\right]\right)\\
& \qquad\times \prod_{i=1}^N\left\<\exp\left(\int d^dx\left(J_i(x)+ \fk(x)\right)\epsilon_i(x) + K_i(x)\cO_i(x)\right)\right\>_0\,.
\end{split}\end{equation}
Here we have used the fact that each Ising model is decoupled when $\lk = \mu = 0$ to factor the correlator. Each expectation value in \eqref{gen2} is then the generating functional for correlators of $\epsilon(x)$ and $\cO(x)$ in a single Ising model:
\begin{equation}\begin{split}
Z_{\text{Ising}}[J_i+\fk,K_i] &= \left\<\exp\left(\int d^dx\left(J_i(x)+ \fk(x)\right)\epsilon_i(x) + K_i(x)\cO_i(x)\right)\right\>_0 \,.
\end{split}\end{equation}
It will prove more convenient, however, to work with the free-energy
\begin{equation}
F_{\text{Ising}}[J_i+\fk,K_i] = -\log Z_{\text{Ising}}[J_i+\fk,K_i]\,,
\end{equation}
which is the generator of connected correlators. We therefore write
\begin{equation}\label{gen3}\begin{split}
Z_{\lk,\mu}[J_i,K_i] &\propto \int D\fk \exp\left(\frac1{\lk}\int d^dx\,\left(\frac1{2}\fk(x)^2 +\mu\fk(x)\right)- \sum_{i=1}^NF_{\text{Ising}}[J_i + \fk,K_i]\right)\,.
\end{split}\end{equation}

The operators $\epsilon_i(x)$ form a reducible representation of the symmetric group $S_N$, splitting into the two representations
\begin{equation}
\fcy E(x) = \sum_{i = 1}^N \epsilon_i(x)\,,\qquad \boldsymbol \epsilon_i(x) = \epsilon_i(x) - \frac 1N\sum_{i = 1}^N \epsilon_i(x)\,.
\end{equation}
We will find it useful to rewrite the sources $J_i(x)$ to match this decomposition, defining
\begin{equation}
\fcy J(x) = \sum_{i = 1}^N J_i(x)\,,\qquad {\bf J}_i(x) = J_i(x) - \frac 1 N \sum_{i = 1}^N J_i(x)\,,
\end{equation}
so that the generating function becomes
\begin{equation}\label{gen4}\begin{split}
Z_{\lk,\mu}&[\fcy J, {\bf J}_i,K_i] \\
&\propto \int D\fk \exp\left(\frac1{\lk}\int d^dx\,\left(\frac1{2}\fk(x)^2 +\mu\fk(x)\right) - \sum_{i=1}^NF_{\text{Ising}}\left[{\bf J}_i + \frac 1N\fcy J + \fk,K_i\right]\right)\,.
\end{split}\end{equation}
Performing the change of variables $\fk\rightarrow\fk - \frac1N\fcy J$, we find
\begin{equation}\label{gen5}\begin{split}
Z_{\lk,\mu} &[\fcy J, {\bf J}_i,K_i]\propto \int D\fk \exp\Bigg(- \sum_{i=1}^NF_{\text{Ising}}\left[{\bf J}_i + \fk,K_i\right]\\
&\qquad +\frac1{\lk}\int d^dx\,\left(\frac12\fk(x)^2  - \frac1{N} \fk(x)\fcy J(x) + \mu\fk(x) + \frac1{2N^2}\fcy J(x)^2 -\frac{\mu}{N}\fcy J(x) \right)\Bigg)\,.\\
\end{split}\end{equation}
The quadratic term in $\fcy J(x)$ does not contribute to correlation functions at separated points, and so we can ignore it. We then see that correlators of $\fk(x)$ are related to those of $\fcy E(x)$ by the relationship
\begin{equation}\label{fieldRedef}
\fcy E(x) = -\frac1{N\lk}\fk(x) - \frac{\mu}{N\lk}\,,
\end{equation}
and so for simplicity we shall redefine our path integral so that $\fcy J(x)$ is the source term for $\fk(x)$ rather than $\fcy E(x)$, defining:
\begin{equation}\begin{split}\label{cubicGen}
&\hat Z_{\lk,\mu}[\fcy J, {\bf J}_i,K_i] \\
&= \int D\fk\, \exp\left(\int d^dx\, \left(\fcy J(x)\fk(x)+\frac{1}{2\lk^2}\fk(x)^2+\frac{\mu}{\lk}\fk(x)\right)- \sum_{i=1}^NF_{\text{Ising}}[{\bf J}_i + \fk,K_i] \right)\,.
\end{split}\end{equation}
We can now take $\lk\rightarrow\infty$ while keeping $\tilde\mu = \frac\mu\lk$ fixed, and so arrive at the expression
\begin{equation}\label{cubicGen2}
Z_{\tilde\mu}[\fcy J, {\bf J}_i,K_i] = \int D\fk\, \exp\left(\int d^dx\, \left(\fcy J(x)\fk(x)+\tilde\mu\fk(x)\right)- \sum_{i=1}^NF_{\text{Ising}}[{\bf J}_i + \fk,K_i]\right)\,.
\end{equation}
Now $\tilde\mu$ is the only dimensionful parameter left. When it is suitably fine-tuned to criticality we can compute correlators of $\fcy E(x)$, $\boldsymbol\epsilon_i(x)$, and $\cO_i(x)$ at the cubic fixed point.

Nothing we have done so far has required $N$ to be large. In order to actually compute \eqref{cubicGen} however, we will need to consider the large $N$ limit. For simplicity, let us first set ${\bf J}_i = K_i = 0$ and so focus only on correlators of $\fk(x)$. To compute these we expand the Ising model free-energy as a sum of connected correlation functions
\begin{equation}
F_{\text{Ising}}[J,0] = -\sum_{k = 1}^\infty \frac{1}{k!} \int d^dx_1\dots d^dx_k\, J(x_1)\dots J(x_k)\<\epsilon(x_1)\dots\epsilon(x_k)\>_{\text{conn}}\,.
\end{equation}
Because the Ising model is conformally invariant, the first few connected correlators take a simple form:
\begin{equation}\begin{split}
\<\epsilon(x)\>_{\text{conn}} &= \<\epsilon(x)\> = 0\,, \\ 
\<\epsilon(x_1)\epsilon(x_2)\>_{\text{conn}}  &= \<\epsilon(x_1)\epsilon(x_2)\> = \frac1{|x_1-x_2|^{2\Dk_\epsilon}}\,,  \\
\<\epsilon(x_1)\epsilon(x_2)\epsilon(x_3)\>_{\text{conn}}  &= \<\epsilon(x_1)\epsilon(x_2)\epsilon(x_3)\> = \frac{f_{\epsilon\epsilon\epsilon}}{|x_1-x_2|^{\Dk_\epsilon}|x_1-x_3|^{\Dk_\epsilon}|x_2-x_2|^{\Dk_\epsilon}}\,,
\end{split}\end{equation}
while the four-point function can ne computed numerically using conformal bootstrap results. We can thus expand
\begin{equation}\label{pertExp}\begin{split}
Z[\fcy J] = \int D\phi&(x) \exp\Bigg(\int d^dx\,\left(\fk(x)\fcy J(x)+\tilde\mu\fk(x)\right)+\frac N2\int d^dx\,d^3y \frac{\fk(x)\fk(y)}{|x-y|^{2\Dk_\epsilon}} \\
 &+N\sum_{k=3}^\infty \frac1{k!}\int d^dx_1\dots d^dx_k\, \fk(x_1)\dots \fk(x_k)\<\epsilon(x_1)\dots\epsilon(x_k)\>_{\text{conn}} \Bigg)\,.
\end{split}\end{equation}

At large $N$ the path-integral integral is dominated by the quadratic term, so that as $N\rightarrow\infty$ the correlators of $\fk(x)$ are given by those of a mean-field theory. For this theory to be conformally invariant we must set the dimensionful parameter $\tilde\mu$ to zero. Computing the $\fk(x)$ two-point function is then straightforward (see for instance \cite{Diab:2016spb}):
\begin{equation}\label{prop}\begin{split}
 \<\fk(x)\fk(y)\> &= \frac{A_d(\Dk_\epsilon)}{N|x-y|^{2d-2\Dk_\epsilon}}\,,\\
\text{ with } \quad 
A_d(\Dk) &= -\frac{\pi^d\Gk(\Dk)\Gk(d-\Dk)}{\Gk\left(\Dk-\frac d2\right)\Gk\left(\frac d2-\Dk\right)}\,,
\end{split}\end{equation}
As expected for a double-trace deformation, $\fk(x)$ has conformal dimension $d-\Dk_\epsilon$ \cite{Stanley:1968gx,PhysRev.176.257,PhysRevLett.31.1494}. This should be compared to $\fcy E(x)$ in the UV theory, which has dimension $\Dk_\epsilon$.

Large $N$ corrections are generated by the higher degree terms in \eqref{pertExp}. We can compute them using a Feynman diagram expansion. For each $k\geq3$ we have a $k$-point vertex, whose value is given by the $k$-point connected correlator in the Ising model:
\begin{equation}\label{kVertN}
\feynmandiagram [large, inline=(b.base), horizontal = b to c] {
a [particle=$z_k$]-- b -- [opacity = 0] c [particle=$\vdots$],
d [particle=$z_3$]-- b -- e [particle=$z_2$],
d-- [opacity = 0] c,
a-- [opacity = 0] c,
f [particle=$z_1$]-- b 
}; = N\<\epsilon(z_1)\dots\epsilon(z_k)\>_{\text{conn}}\,.
\end{equation}
Each of these vertices corresponds to one term of the infinite sum in the second line of \eqref{pertExp}; we can think of each as a non-local $\fk^k$ interaction. In a Feynman diagram each leg of a vertex is either connected to another leg, or it is connected to an external operator. Internal legs are connected to other legs via the propagator
\begin{equation}\label{intLeg}
\feynmandiagram [inline=(b.base), horizontal = a to b] {
a [particle=$z_1$] --  b  [particle=$z_2$]}; \equiv \<\fk(z_1)\fk(z_2)\>_{N=\infty} =  \frac{A_d(\Dk_\epsilon)}{N|x-y|^{2d-2\Dk_\epsilon}}\,,
\end{equation}
and we then integrate over both $z_1$ and $z_2$. Legs connected to external operators take a similar form:
\begin{equation}\label{extLeg}
\feynmandiagram [inline=(b.base), horizontal = a to b] {
a [particle=$\fk(x)$] --  b  [particle=$z$]}; \equiv \<\fk(x)\fk(z)\>_{N=\infty} =  \frac{A_d(\Dk_\epsilon)}{N|x-z|^{2d-2\Dk_\epsilon}}\,,
\end{equation}
but this time we only integrate over $z$, but not $x$. Finally, we can always connected two external operators to each other:
\begin{equation}\label{extLeg2}
\feynmandiagram [inline=(b.base), horizontal = a to b] {
a [particle=$\fk(x)$] --  b  [particle=$\fk(y)$]}; \equiv \<\fk(x)\fk(y)\>_{N=\infty} =  \frac{A_d(\Dk_\epsilon)}{N|x-y|^{2d-2\Dk_\epsilon}}\,.
\end{equation}

To compute an $n$-point correlator $\<\fk(x_1)\dots\fk(x_n)\>$, we sum over all Feynman diagrams which have an external leg $\fk(x_i)$ for each $i=1\,,\dots\,,n$. As in any Feynman diagram expansion, we must divide by the symmetry factor; this purely combinatorial factor is identical to that found in a local theory with $\fk^k$ interactions. Counting the overall order of a diagram is straightforward: each vertex contributes a power of $N$, while each propagator gives a power of $N^{-1}$.

As an example, the leading corrections to the two-point function are given by:
\begin{equation}\begin{split}\label{2ptDiag}
\<\fk(x)\fk(y)\>_{\text{conn}} &= \feynmandiagram [inline=(b.base), horizontal = a to b] {
a [particle=$\fk(x)$] --  b  [particle=$\fk(y)$]}; + \feynmandiagram [inline=(b.base), horizontal = a to d] {
a [particle=$\fk(x)$] --  b  -- [half left]  c --  d [particle=$\fk(y)$],
c -- [half left ] b
}; \\
&+ \feynmandiagram [inline=(b.base), horizontal = a to d] {
a [particle=$\fk(x)$] --  b  -- [half left]  c -- [half left]   b -- d [particle=$\fk(y)$]
}; + O(N^{-3})\,,
\end{split}\end{equation}
where we have dropped any disconnected and tadpole diagram which may occur at $O(N^{-1})$. To illustrate our rules let us write the first diagram in \eqref{2ptDiag} explicitly in terms of Ising model correlators. Rewriting the diagram to make the internal spacetime points explicit, we see that:
\begin{equation}\begin{split}\label{2ptDiagv2}
&\feynmandiagram [inline=(b.base), horizontal = a to d] {
a [particle=$\fk(x)$] --  b  -- [half left]  c --  d [particle=$\fk(y)$],
c -- [half left ] b
};  = \feynmandiagram [inline=(b.base), horizontal = a to h] {
a [particle=$\fk(x)$] -- m --  b [particle=$z_1$]  -- c --  d [particle=$z_2$] -- p -- e [particle = $z_4$] -- f -- g [particle=$z_6$] -- n -- h [particle=$\fk(y)$] ,
c --  i [particle=$z_3$] -- q -- j [particle = $z_5$] -- f,
p -- [draw=none] q,
d -- [draw=none] i,
e -- [draw=none] j
}; \\
&\qquad\qquad = \frac{A_d(\Dk_\epsilon)^4}{2N^2} \int \frac{d^dz_1\dots d^dz_6}{\left(z_{24}z_{35}|x-z_1| |y-z_6|\right)^{2d-2\Dk_\epsilon}}\<\epsilon(z_1)\epsilon(z_2)\epsilon(z_3)\>\<\epsilon(z_4)\epsilon(z_5)\epsilon(z_6)\> \\
&\qquad\qquad = \frac{f_{\epsilon\epsilon\epsilon}^2A_d(\Dk_\epsilon)^4}{2N^2} \int \frac{d^dz_1\dots d^dz_6}{\left(z_{24}z_{35}|x-z_1| |y-z_6|\right)^{2d-2\Dk_\epsilon}\left(z_{12}z_{13}z_{23}z_{45}z_{46}z_{56}\right)^{\Dk_\epsilon}}\,.
\end{split}\end{equation}
We will evaluate this integral in the next section, along with the other diagram \eqref{2ptDiag}. Together, they allow us to compute the leading correction to the anomalous dimension of $\fk$ at large $N$.

Let us next turn to correlators of $\cO_i(x)$. For simplicity we begin with the 1pt function, which we can compute by differentiating the generating functional:
\begin{equation}\begin{split}\label{sigmaCor}
\<\cO_{i}(x)\> &= -\frac1{Z[0,0,0]}\int D\fk\, \frac{\dk F_{\text{Ising}}[\fk,K]}{\dk K(x)}\Bigg|_{K = 0}\exp\left(- \sum_{j=1}^NF_{\text{Ising}}[\fk,0]\right) \,.
\end{split}\end{equation}
We can then expand the derivative of the free energy as a sum of connected correlators
\begin{equation}
-\frac{\dk F_{\text{Ising}}[\fk,K]}{\dk K(x)}\Bigg|_{K = 0} = \sum_{k = 0}^\infty \frac 1 {k!}\int \left(\prod_{i=1}^k dz_i\,\fk(z_i)\right)\left\<\cO(x)\prod_{i=1}^k \epsilon(z_i)\right\>_{\text{conn}}\,,
\end{equation}
and so find that
\begin{equation}\begin{split}\label{sigSeries}
\<\cO_{i}(x)\> &= \left\<\frac{\dk F_{\text{Ising}}[\fk,K]}{\dk K(x)}\Bigg|_{K = 0}\right\>_{\text{cubic}}\\
&= \sum_{k = 0}^\infty \frac1{k!} \int dz_1\dots dz_k\,\left\<\fk(z_1)\dots\fk(z_k)\right\>_{\text{cubic}}\left\<\cO(x)\epsilon(z_1)\dots\epsilon(z_k)\right\>_{\text{conn}}\,.
\end{split}\end{equation}
The subscripts on the correlators denote that the expectation values of $\fk(z)$ are taken at the cubic fixed point, while the correlators of $\cO(x)$ and $\epsilon(x)$ are connected correlators in a single Ising model. We have already seen how to compute the $\fk(z)$ correlators using regular Feynman diagrams. To represent \eqref{sigSeries} diagrammatically, we must simply introduce new vertices containing an external $\cO(x)$ operator along with $k$ internal $\fk(z)$ lines, whose value is the connected correlator $\<\cO(x)\epsilon(z_1)\dots\epsilon(z_k)\>$:
\begin{equation}
\feynmandiagram [small, inline=(b.base), horizontal = a to b] {
a [particle=$\cO_i(x)$] -- [scalar] b ,
b  --  d [particle=$z_1$] -- [opacity = 0] e [particle=$\vdots$] ,
b  --  e ,
b  --  f [particle=$z_k$] -- [opacity = 0] e 
}; = \<\cO(x)\epsilon(z_1)\dots\epsilon(z_k)\>_{\text{conn}}\,.
\end{equation}
We can then write the first few terms in \eqref{sigSeries} at large $N$ as
\begin{equation}\begin{split}
\<\cO_{i}(x)\> &= \feynmandiagram [inline=(b.base), horizontal = a to b] {
a [particle=$\cO_i(x)$] -- [scalar] b ,
b  -- [half left] d -- [half left] b  
}; + \feynmandiagram [inline=(b.base), horizontal = a to b] {
a [particle=$\cO_i(x)$] -- [scalar] b ,
b  -- [quarter left] d -- [quarter left] b  ,
b  -- [quarter left] e -- [quarter left] b  
}; + \feynmandiagram [inline=(b.base), horizontal = a to b] { 
a [particle=$\cO_i(x)$] -- [scalar] b ,
b  -- [half left] d ,
b  -- d,
b  -- [half right] d  
}; \\
&\feynmandiagram [inline=(b.base), horizontal = a to b] {
a [particle=$\cO_i(x)$] -- [scalar] b ,
c -- b -- d,
d  --  c -- [half left] d  
};+ O(N^{-3})\,.
\end{split}\end{equation}
In practice we can always fix $\<\cO_i(x)\> = 0$ through a field redefinition, and so can automatically subtract of these tadpole diagrams.

Next consider the higher point correlator
\begin{equation}\label{HSSigma}\begin{split}
\<\cO_{i_1}(x_1)\dots&\cO_{i_n}(x_n)\> = \frac1{Z}\left.\frac{\dk}{\dk K_{i_1}(x_1)}\dots\frac{\dk}{\dk K_{i_n}(x_n)}Z[0, 0,K_j]\right|_{K_j=0} \\
&= \frac1{Z}\int D\fk\, \left.\frac{\dk}{\dk K_{i_1}(x_1)}\dots\frac{\dk}{\dk K_{i_n}(x_n)}\exp\left(- \sum_{j=1}^NF_{\text{Ising}}[\fk,K_j]\right)\right|_{K_j = 0}\,.
\end{split}\end{equation}
If all of the $i_k$'s are distinct, each derivative acts on a different copy of the Ising model and so
\begin{equation}\begin{split}
\<\cO_{i_1}(x_1)\dots&\cO_{i_n}(x_n)\> = (-1)^n\left\<\prod_{j = 1}^n\left.\frac{\dk F[\fk,K]}{\dk K(x_j)}\right|_{K = 0}\right\>_{\text{cubic}} \\
&= \sum\text{diagrams with each }\sk_{i_k}(x_k)\text{ leg connected to a different vertex}\,.
\end{split}\end{equation}
As an example, the first few contributions to the two-point function $\<\cO_1(x_1)\cO_2(x_2)\>_{\text{conn}}$ are
\begin{equation}\begin{split}
\<\cO_1(x_1)\cO_2(x_2)\>_{\text{conn}} &=  \feynmandiagram [inline=(b.base), horizontal = a to b] {
c [particle=$\cO_1(x_1)$] -- [scalar] a -- [half left] b -- [half left] a ,
b  -- [scalar] d [particle=$\cO_2(x_2)$],
a --[draw=none] b,
a --[draw=none] e --[draw=none] b
};  + \feynmandiagram [inline=(b.base), horizontal = a to b] {
i1 [particle=$\cO_1(x_1)$] -- [scalar] a --  [half left ] b -- [half left ] a ,
b  -- [scalar] i2 [particle=$\cO_2(x_2)$] ,
a -- b,
a --[draw=none] e --[draw=none] b
}; \\
&+ \feynmandiagram [inline=(b.base), horizontal = a to b] {
i1 [particle=$\cO_1(x_1)$] -- [scalar] a -- [quarter left]  b  -- [scalar] i2 [particle=$\cO_2(x_2)$],
a -- [quarter right] c -- [quarter right] b,
c -- [quarter left] b
}; + \feynmandiagram [inline=(b.base), horizontal = a to b] {
i1 [particle=$\cO_1(x_1)$] -- [scalar] a -- [quarter left]  b  -- [scalar] i2 [particle=$\cO_2(x_2)$],
a -- [quarter right] c -- [quarter right] b,
a -- [quarter left] c
};\\
&+ O(N^{-4})\,.
\end{split}\end{equation}

If multiple $\cO_i(x)$ operators have the same index $i$, additional terms will appear in \eqref{HSSigma} where a single Ising model free energy is differentiated multiple times. For instance,
\begin{equation}
\<\cO_1(x_1)\cO_1(x_2)\> = \left\<\left.\frac{\dk F}{\dk K(x_1)}\frac{\dk F}{\dk K(x_2)}\right|_{K=0}\right\> - \left\<\left.\frac{\dk^2 F}{\dk K(x_1)\dk K(x_2)}\right|_{K=0} \right\> \,.
\end{equation}
We have already seen how to compute the first term. To compute the second, we note that
\begin{equation}
-\left.\frac{\dk^2 F}{\dk K(x_1)\dk K(x_2)}\right|_{K=0} = \sum_{k = 0}^\infty \frac1{k!}\int\left(\prod_{j=1}^k dz_j\,\fk(z_k)\right)\left\<\sk(x_1)\sk(x_2)\epsilon(z_1)\dots\epsilon(z_k)\right\>_{\text{conn}}\,.
\end{equation}
To represent these, we introduce vertices with two external lines, along with $k$ internal $\fk(z)$ lines, whose value is the connected correlator $\<\cO(x_1)\cO(x_n)\epsilon(z_1)\dots\epsilon(z_k)\>$.  For instance, when $k = 2$ we have
\begin{equation}
\feynmandiagram [inline=(b.base), horizontal = a to d] {
a [particle=$\cO_i(x_1)$] --  [scalar] b  -- c [particle=$z_1$],
d [particle=$z_2$]      --  b  --   [scalar] e [particle=$\cO_j(x_1)$]
}; = \delta_{ij}\<\sigma(x_1)\sigma(x_2)\epsilon(z_1)\epsilon(z_2)\>_{\text{conn}}\,.
\end{equation}
Because of the Kronecker delta $\delta_{ij}$, this vertex only appears if $i = j$, so that both $\cO$s belong to the same copy of the Ising model. With this addition, we can write the first few corrections to the two-point function $\<\cO_1(x_1)\cO_1(x_2)\>$:
\begin{equation}\label{sgsgTerm}\begin{split}
\<\cO_1(x_1)&\cO_1(x_2)\>_{\text{conn}} - \<\cO_1(x_1)\cO_2(x_1)\>_{\text{conn}} \\
&=
\feynmandiagram [inline=(b.base), horizontal = a to c] {
a [particle=$\cO_1(x_1)$] --  [scalar] b [dot] -- [scalar] c [particle=$\cO_1(x_2)$],
b -- [quarter left,draw=none] d -- [quarter left,draw=none] b,
b -- [quarter left,draw=none] e -- [quarter left,draw=none] b
};+
\feynmandiagram [inline=(b.base), horizontal = a to c] {
a [particle=$\cO_1(x_1)$] --  [scalar] b  -- [scalar] c [particle=$\cO_1(x_2)$],
b -- [half left] d -- [half left] b
};\\
& + 
\feynmandiagram [inline=(b.base), horizontal = a to c] {
a [particle=$\cO_1(x_1)$] --  [scalar] b  -- [scalar] c [particle=$\cO_1(x_2)$],
b -- [quarter left] d -- [quarter left] b,
b -- [quarter left] e -- [quarter left] b
}; + 
\feynmandiagram [inline=(b.base), horizontal = a to c] {
a [particle=$\cO_1(x_1)$] --  [scalar] b  -- [scalar] c [particle=$\cO_1(x_2)$],
b -- [half left] d -- [half left] b,
b -- d
}; + O(N^{-3})\,,
\end{split}\end{equation}
where we have dropped disconnected and tadpole contributions. Note that although the first diagram in \eqref{sgsgTerm} may look superficially like a propagator for $\cO_i(x)$, strictly speaking it is a vertex with two external legs and no internal legs. Unlike the $\phi$ propagator, which occur internally in diagrams connecting vertices together, dashed lines only ever appear as external legs for diagrams.

Extending our diagrams to describe correlators where more than two $\cO_i(x)$ operators belong to the same copy of the Ising model is straightforward. We simply need to introduce vertices with arbitrary numbers of external $\cO_i(x)$ legs:
\begin{equation}
\feynmandiagram [large,inline=(b.base), horizontal = a to d] {
p [particle=$\vdots$] -- [opacity=0] b -- q [particle=$z_1$] ,
r -- [opacity=0] b -- [opacity=0] s [particle=$\vdots$] ,
a [particle=$\cO_{i_m}(x_m)$] --  [scalar] b  -- [opacity=0] c ,
d [particle=$z_k$]      --  b  --   [scalar] e [particle=$\cO_{i_1}(x_1)$]
}; = \delta_{i_1i_2\dots i_m} \left\<\cO(x_1)\dots\cO(x_m)\epsilon(z_1)\dots\epsilon(z_k)\right\>\,,
\end{equation}
where $\delta_{i_1i_2\dots i_m}$ is the rank-$m$ tensor which is one if all indices have the same value, and zero otherwise. 
To compute correlators we then simply sum over all possible contributing diagrams with the correct external legs.

So far we have let $\cO_i(x)$ be any arbitrary operator in the 3d Ising model (other than the $\epsilon_i(x)$ operators used to define the double-trace deformation). We are most interested, however, in the special case of $\sk_i(x)$. Due to the $\mathbb Z_2$ symmetry of each individual Ising model, any vertex with an odd number of $\sk_i(x)$ external legs vanishes, reducing the number of diagrams we have to compute. Likewise, any correlator with an odd number of $\sk_i(x)$'s (for any specific choice of $i = 1\,,\dots\,,N$) will automatically vanish due to hypercubic symmetry. 

Consider now more general correlators
\begin{equation}
\<\cO^{a_1}_{i_1}(x_1)\dots\cO^{a_n}_{i_n}(x_n)\>\,,
\end{equation}
where each $\cO^a(x)$ for $a=1\,,2\,,\dots$ is a different kind of local operator in the 3d Ising model (again excluding the $\epsilon(x)$ operator, which we will discuss shortly). So for instance, we might want to consider correlators including both the $\sigma_i(x)$ operators and the stress-tensor $T_i^{\mu\nu}(x)$ operators. Generalizing our diagrams to this case is straightforward: we simply introduce vertices with more general external legs
\begin{equation}
\feynmandiagram [large,inline=(b.base), horizontal = a to d] {
p [particle=$\vdots$] -- [opacity=0] b -- q [particle=$z_1$] ,
r -- [opacity=0] b -- [opacity=0] s [particle=$\vdots$] ,
a [particle=$\cO^{a_m}_{i_m}(x_m)$] --  [scalar] b  -- [opacity=0] c ,
d [particle=$z_k$]      --  b  --   [scalar] e [particle=$\cO^{a_1}_{i_1}(x_1)$]
}; = \delta_{i_1\dots i_m} \left\<\cO^{a_1}(x_1)\dots\cO^{a_m}(x_m)\epsilon(z_1)\dots\epsilon(z_k)\right\>\,.
\end{equation}
We compute correlators by summing over all diagrams with the appropriate external legs.

Finally, let us turn to correlators of ${\boldsymbol \epsilon}_i(x)$, which we can compute by differentiating with respect to ${\bf J}_i$. Even though these satisfy the constraint $\sum_i {\bf J}_i = 0$, we can treat the sources independently because
\begin{equation}\begin{split}
\sum_{i = 1}^N \frac{\dk}{\dk {\bf J}_i(x)} Z[\fcy J,{\bf J},K] 
&= \int D\fk\, \sum_{i = 1}^N \frac{\dk}{\dk {\bf J}_i(x)}e^{- \sum F_{\text{Ising}}[{\bf J}_i + \fk,K_i] + \int d^dx\, \fcy J(x)\fk(x)} \\
&=  \int D\fk\, \left(\frac{\dk}{\dk \fk(x)}-\fcy J(x)\right)e^{- \sum F_{\text{Ising}}[{\bf J}_i + \fk,K_i] + \int d^dx\, \fcy J(x)\fk(x)} \\
&= -\fcy J(x) Z[\fcy J,{\bf J},K]\,,
\end{split}\end{equation}
so that, up to contact terms, the operator $\sum_i {\boldsymbol \epsilon}_i(x)$ automatically vanishes. We can therefore treat the ${\boldsymbol\epsilon}_i(x)$ akin to other local operators, introducing external vertices such as
\begin{equation}
\feynmandiagram [large,inline=(b.base), horizontal = a to d] {
p [particle=$z_k$] --  b -- [opacity = 0] q  [particle=$\vdots$],
r [particle=$\epsilon_{j_n}(y_n)$] --  [scalar] b -- [scalar] s [particle=$\cO_{i_m}(x_m)$] ,
a [particle=$\epsilon_{j_1}(y_1)$] --  [scalar] b  -- [opacity = 0]  c [particle=$\ddots$] ,
d [particle=$\dots$]  -- [opacity = 0]  b  --   e [particle=$z_1$] ,
v1 [particle=$\cO_{i_1}(x_1)$] -- [scalar] b 
}; = \delta_{i_1\dots i_mj_1\dots j_n} \left\<\cO(x_1)\dots\cO(x_m)\epsilon(y_1)\dots\epsilon(y_n)\epsilon(z_1)\dots\epsilon(z_k)\right\>\,.
\end{equation}
 Correlators are again computed by summing over all diagrams with the appropriate external legs.

So far we have avoided the issue of divergences. In general, we should expect diagrams to suffer from both power-law and logarithmic divergences. The power-law divergences occur because of the $\tilde\mu\fk(x)$ term in \eqref{pertExp}, which is a relevant interaction, and can be subtracted through an (infinite) shift of $\tilde\mu$. Logarithmic divergence, on the other hand, appear because of the possibility of operator redefinitions $\fcy O\rightarrow Z_{\fcy O}\fcy O$, and can be removed by (infinite) field redefinitions. 

In this paper we will compute leading corrections to anomalous dimension, and so we will only need to compute one-loop diagrams. Because of this, we can subtract off divergences in an \emph{ad hoc} fashion. Higher-loop calculations however require a more careful treatment. A common strategy used to study the large $N$ limit of the $O(N)$ models is to introduce a long-ranged kinetic term for the auxiliary field \cite{Vasiliev:1981dg,Vasiliev:1981yc,Vasiliev:1975mq}, a technique we expect should generalize without difficulty to the case considered here.

\section{Anomalous Dimensions}
\label{ANOM}

Having derive expressions for cubic correlators at large $N$ expansion, we will now evaluate the leading corrections to conformal dimensions of operators at the cubic fixed point.

\subsection{Single-trace operators}

\subsubsection{$\sigma$}
\label{SIGANOM}

The $\sigma_i(x)$ operators form a single irreducible representation of the $S_N\ltimes\mathbb Z_2^N$ hypercubic symmetry group. When suitably normalized, their two-point function is
\begin{equation}\label{largeNSig}
\<\sigma_i(x)\sigma_j(0)\> = \frac{\delta_{ij}}{|x|^{2\Dk_\sk^*(N)}}\,,
\end{equation}
where $\Dk^*_\sk(N)$ is the conformal dimension of $\sigma_i(x)$ at the cubic fixed point. Using \eqref{sgsgTerm} and noting that $\<\sk_i(x)\sk_j(x)\>$ automatically vanishes due to hypercubic symmetry, we find that
\begin{equation}
\<\sigma_i(x)\sigma_j(0)\> = \delta_{ij}\left(\frac{1}{|x|^{2\Dk_\sk}} + \feynmandiagram [inline=(b.base), horizontal = a to c] {
a [particle=$\sk(x)$] --  [scalar] b  -- [scalar] c [particle=$\sk(0)$],
b -- [half left] d -- [half left] b
}; + O(N^{-2})\right)\,,
\end{equation}
and so are left with a single diagram to compute:
\begin{equation}\label{PiSigSig}
\Pi_{\sk\sk}(x) = \feynmandiagram [inline=(b.base), horizontal = a to c] {
a [particle=$\sk(x)$] --  [scalar] b  -- [scalar] c [particle=$\sk(0)$],
b -- [half left] d -- [half left] b
}; = \frac{A_d(\Dk_\epsilon)}{2N}\int \frac{d^dz_1d^dz_2}{|z_1-z_2|^{2d-2\Dk_\epsilon}}\<\sk(x)\sk(0)\epsilon(z_1)\epsilon(z_2)\>_{\text{conn}}\,.
\end{equation}
The integral diverges as both $z_1$ and $z_2$ approach either $x$ or $0$. To regulate it, we cut out little spheres of size $\Lambda$ around both $x$ and the origin when performing the $z_1$ integral. We then note that under conformal inversion, 
\begin{equation}
\<\sk(x)\sk(0)\epsilon(z_1)\epsilon(z_2)\>_{\text{conn}} = \frac{\left\<\sk\left(\frac{x}{|x|^2}\right)\sk(\infty)\sk\left(\frac{z_1}{|z_2|^2}\right)\sk\left(\frac{z_1}{|z_2|^2}\right)\right\>_{\text{conn}}}{|x|^{2\Dk_\sk}(|z_1||z_2|)^{2\Dk_\epsilon}}\,.
\end{equation}
Performing the change of variables
\begin{equation}
z_i^\mu \rightarrow \frac{z_i^\mu}{|z_i|^2}\,,
\end{equation}
we thus find that
\begin{equation}
\Pi_{\sk\sk}(x;\Lk) = \frac 1 {|x|^{2\Dk_\sk}}\frac{A_d(\Dk_\epsilon)}{N}\int_{\fcy C}\frac{d^dz_1d^dz_2}{|z_1-z_2|^{2d-2\Dk_\epsilon}}\left\<\sk\left(\frac{x}{|x|^2}\right)\sk(\infty)\epsilon(z_1)\epsilon(z_2)\right\>_{\text{conn}}\,,
\end{equation}
where we integrate over the region $\fcy C$ for which $z_1$ satisfies
\begin{equation}
|z_1|\leq \Lk^{-1}\quad \text{ and }\quad \left|z_1-\frac{x}{|x|^2}\right|\geq \frac{|z_1|\Lk}{|x|}\,.
\end{equation}
As we take $\Lk\rightarrow0$, however, we can replace the latter condition with the simpler one
\begin{equation}
 \left|z_1-\frac{x}{|x|^2}\right|\geq \frac{\Lk}{|x|^2}\,,
\end{equation}
which is equivalent to the original condition up to $O(\Lk)$ corrections. We then perform a second change of variables $z_i \rightarrow z_i - \frac{x}{|x|^2}$ to derive
\begin{equation}
\Pi_{\sk\sk}(x) = \frac 1 {|x|^{2\Dk_\sk}}\frac{A_d(\Dk_\epsilon)}{2N}\int_{|z_1|>\Lk|x|^{-2}}^{|z_1|<\Lk^{-1}}\frac{d^dz_1d^dz_2}{|z_1-z_2|^{2d-2\Dk_\epsilon}}\left\<\sk(0)\sk(\infty)\epsilon(z_1)\epsilon(z_2)\right\>_{\text{conn}}\,.
\end{equation}
Now observe that the integral is rotationally invariant, so that we can set $z_1 = r\hat e_1$ without loss of generality, where $\hat e_1$ is the unit vector $(1,0,\dots,0)$. Performing a final change of variables $z_2\rightarrow r w$ and then using scale invariance, we can write
\begin{equation}\begin{split}
\Pi_{\sk\sk}(x;\Lk) &= \frac {A_d(\Dk_\epsilon)S_d}{2N|x|^{2\Dk_\sk}}\int_{r>\Lk|x|^{-2}}^{r<\Lk^{-1}}\frac{dr}r\int\frac{d^dw}{|w-\hat e_1|^{2d-2\Dk_\epsilon}}\left\<\sk(0)\sk(\infty)\epsilon(\hat e_1)\epsilon(w)\right\>_{\text{conn}} \\ 
&= \frac {A_d(\Dk_\epsilon)S_d\log\left(|x|^2\Lk^{-2}\right)}{2N|x|^{2\Dk_\sk}}\int\frac{d^dw}{|w-\hat e_1|^{2d-2\Dk_\epsilon}}\left\<\sk(0)\sk(\infty)\epsilon(\hat e_1)\epsilon(w)\right\>_{\text{conn}} \,,
\end{split}\end{equation}
where $S_d=\frac{2\pi^{d/2}}{\Gk(d/2)}$ is the surface area of the unit sphere in $d$ dimensions. Defining 
\begin{equation}\label{zetaSig}
\zeta_\sigma = S_d\int\frac{d^dw}{|w-\hat e_1|^{2d-2\Dk_\epsilon}}\left\<\sk(0)\sk(\infty)\epsilon(\hat e_1)\epsilon(w)\right\>_{\text{conn}}\,,
\end{equation}
and subtracting off the logarithmic divergence which appears as $\Lk\rightarrow0$, we arrive at the expression
\begin{equation}\begin{split}
\Pi_{\sk\sk}(x) &= \frac {A_d(\Dk_\epsilon)\zeta_\sigma\log\left(|x|^2\right)}{2N|x|^{2\Dk_\sk}}\,.
\end{split}\end{equation}
Comparing this correction to the exact result \eqref{largeNSig} for the two-point function, we conclude that
\begin{equation}\label{sigAnom}
\Dk^*_\sk(N) = \Dk_\sk - \frac{A_d(\Dk_\epsilon)\zeta_\sigma}{2N} + O(N^2)\,.
\end{equation}

All that remains for us to compute is $\zeta_\sigma$. Integrals of this form were considered in \cite{Komargodski:2016auf,Behan:2259464}, where they arise in calculations of $\beta$-functions and anomalous dimensions in conformal perturbation theory. Our integral \eqref{zetaSig} suffers from power-law divergences when $w\rightarrow\hat e$, which we must subtract. As shown in \cite{Komargodski:2016auf}, the integral \eqref{zetaSig} can be rewritten as
\begin{equation}\label{pracZeta}
\zeta_\sigma = S_d\int_{\fcy R} d^dw\left(\frac{2\<\sigma(0)\epsilon(w)\epsilon(\hat e)\sigma(\infty)\>_{\text{conn}}}{|w-\hat e|^{2d-2\Dk_\epsilon}} + \frac{\<\epsilon(0)\epsilon(w)\sigma(\hat e)\sigma(\infty)\>_{\text{conn}}}{|w|^{2d-2\Dk_\epsilon}}\right)\,,
\end{equation}
where $\fcy R = \{w\,:\,|w|<1\,, |w|<|w-\hat e|\}$. Only the second term in this expression has a power-law divergence, which appears as $w\rightarrow0$ and must be subtracted. We now expand $\<\sigma\sigma\epsilon\epsilon\>$ as a sum of conformal blocks:
\begin{equation}\begin{split}
\<\sigma(0)\epsilon(w)\epsilon(\hat e)\sigma(\infty)\> &= \frac1{|w|^{\Dk_\sk+\Dk_\epsilon}}\sum_{\fcy O}f^2_{\sk\epsilon\fcy O}g^{\Dk_\epsilon-\Dk_\sk,\Dk_\sk-\Dk_\epsilon}_{\Dk_\cO,\ell_\cO}(z,\bar z)\,, \\
\<\epsilon(0)\epsilon(w)\sigma(\hat e)\sigma(\infty)\> &= \frac1{|w|^{2\Dk_\epsilon}}\sum_{\fcy O}f_{\epsilon\epsilon\fcy O}f_{\sk\sk\fcy O}g^{0,0}_{\Dk_\cO,\ell_\cO}(z,\bar z)\,, \\
\end{split}\end{equation}
where $z = w_1 + i\sqrt{w\cdot w-w_1^2}$, and where $g_{\Dk_\cO,\ell_\cO}(z,\bar z)$ are the conformal blocks computed as in \cite{Kos:2014bka}. The low-lying conformal dimensions and OPE coefficients contributing to these correlators have been computed using the numeric conformal bootstrap \cite{Simmons-Duffin:2016wlq}, and so we can sum the operators with $\Dk<\Dk_* = 12$ in order to approximate these four-point functions. The truncation error is approximately bound by \cite{Pappadopulo:2012jk,Behan:2259464}:
\begin{equation}\label{trunboun}\begin{split}
\left|\sum_{\fcy O}f^2_{\sk\epsilon\fcy O}g^{\Dk_\epsilon-\Dk_\sk,\Dk_\sk-\Dk_\epsilon}_{\Dk_\cO,\ell_\cO}(z,\bar z)\right| &\lesssim \frac{(2\Dk_*)^{2\Dk_\sk+2\Dk_\epsilon}}{\sqrt{\Gk(4\Dk_\sk+1)\Gk(4\Dk_\epsilon+1)}}\frac{|z|}{|1+\sqrt{1-z}|^2} \,, \\
\left|\sum_{\fcy O}f_{\epsilon\epsilon\fcy O}f_{\sk\sk\fcy O}g^{0,0}_{\Dk_\cO,\ell_\cO}(z,\bar z)\right|&\lesssim \frac{(2\Dk_*)^{2\Dk_\sk+2\Dk_\epsilon}}{\Gk(2\Dk_\sk+2\Dk_\epsilon+1)}\frac{|z|}{|1+\sqrt{1-z}|^2}\,.\\
\end{split}\end{equation}
We then apply the method described in Appendix C of \cite{Komargodski:2016auf} to numerically evaluate \eqref{pracZeta} block by block. We find that
\begin{equation}\label{zetaSig2}
\zeta_\sigma \approx -6.40\pm0.02
\end{equation}
for the 3d Ising model, where the error is dominated by the uncertainty in the OPE coefficients computed in \cite{Simmons-Duffin:2016wlq}, rather than from the truncation bound \eqref{trunboun} or from the numerical error in the integration. However, the error estimate should be taken with a grain of salt, as it assumes both that the OPE uncertainties are independent and that no operators are missing from \cite{Simmons-Duffin:2016wlq} with $\Dk<12$. Plugging this into \eqref{sigAnom}, we conclude that
\begin{equation}
\Dk^*_\sk = 0.518149  + \frac{0.000616}{N} + O(N^{-2})\,.
\end{equation}

The leading large $N$ correction to $\Dk^*_\sk$ is much smaller than the typical error estimate for $\Dk_\sk^*$ when computed using either the $4-\epsilon$ or $d=3$ perturbative expansions, and so we cannot usefully compare our results to these calculations.\footnote{A table of the critical exponent $\gamma = \frac{(d-2\Dk_\sk^*)}{(d-\Dk_{\fcy E}^*)}$ values, computed using various resummations of the $4-\epsilon$ and $d=3$ expansions, can be found in \cite{PhysRevB.62.12195}. For $N=\infty$, the estimate with the least uncertainty is $\Dk_{\sk,\infty}^* = 0.516(1)$, which both considerably smaller and much more uncertain then the bootstrap result $\Dk_{\sk,\infty}^* = \Dk_\sk = 0.5181489(10)$. Finite $N$ estimates suffer from similar issues.} The only value of $\Dk^*_\sk$ we have a precise handle on is when $N=3$. For this special case we expect the conformal dimension to be close to that of the $O(3)$ model, with
\begin{equation}
\Dk_{\sk,\ \text{O(3)}} - \Dk^*_{\sk,\ N=3\ \text{cubic}} = 0.00005(5)
\end{equation}
estimated from the $d=3$ perturbative expansion \cite{Folk01}. Combining this with the current best result for $\Dk_{\sk,\ \text{O(3)}} = 0.518920(25)$, computed using Monte Carlo \cite{PhysRevB.102.024406}, we conclude that
\begin{equation}
\Dk^*_{\sk,\ N=3\ \text{cubic}} = 0.51887(6)\,.
\end{equation}
We can compare this to the large $N$ estimate
\begin{equation}
\Dk^*_{\sk,\ N=3\ \text{cubic}} \approx 0.518149  + \frac{0.000616}{3} = 0.51835\,.
\end{equation}

\subsubsection{$\fcy E(x)$ and $\boldsymbol \epsilon_i(x)$}

Our next task is to compute the anomalous dimension of $\boldsymbol\epsilon_i(x)$. To this end, we expand
\begin{equation}\begin{split}\label{ekek2}
\<\boldsymbol\epsilon_1(x)\boldsymbol\epsilon_1(0)\>_{\text{conn}} &= 
\feynmandiagram [inline=(a.base), horizontal = a to c] {
a  [particle=$\epsilon_1(x)$]-- [scalar] b [dot] -- [scalar] c [particle=$\epsilon_1(0)$]
};  + 
\feynmandiagram [inline=(a.base), horizontal = a to d] {
a  [particle=$\epsilon_1(x)$] -- [scalar]  b  --  c -- [scalar] d [particle=$\epsilon_1(0)$]
};  \\
&+ 
\feynmandiagram [inline=(a.base), horizontal = a to c] {
a [particle=$\epsilon(x)$] --  [scalar] b  -- [scalar] c [particle=$\epsilon(0)$],
b -- [half left] d -- [half left] b
}; + O(N^{-2})\,.
\end{split}\end{equation}
The first diagram is an vertex with only two external $\epsilon_1(x)$ legs, which is simply the two-point function of $\epsilon(x)$ in the Ising model,
\begin{equation}
\feynmandiagram [inline=(a.base), horizontal = a to c] {
a  [particle=$\epsilon_1(x)$] -- [scalar] b [dot] -- [scalar] c [particle=$\epsilon_1(0)$]
}; = \<\epsilon(x)\epsilon(0)\> = \frac1{|x|^{2\Dk_\epsilon}}\,.
\end{equation}
The second diagram is given by
\begin{equation}\begin{split}
\feynmandiagram [inline=(a.base), horizontal = a to d] {
a  [particle=$\boldsymbol\epsilon_1(x)$] -- [scalar]  b  --  c -- [scalar] d [particle=$\boldsymbol\epsilon_1(0)$]
}; &= \lim_{\delta\rightarrow0} \frac{A_d(\Dk_\epsilon)}{N} \int \frac{d^dz_1 d^dz_2}{|x-z_1|^{2\Dk_\epsilon}z_{12}^{2d-2\Dk_\epsilon+\dk}|z_2|^{2\Dk_\epsilon}} \\
&= -\frac1{N|x|^{2\Dk_\epsilon}}\,,
\end{split}\end{equation}
where we have introduced $\dk$ as a regulator, evaluated the integrals by using
\begin{equation}\label{2ptInt}
\int \frac{d^dz}{|x-z|^{2\ak}|y-z|^{2\bk}} = \frac{\pi^{d/2}}{|x-y|^{2\ak+2\bk-d}}\frac{\Gk\left(\frac d2-\ak\right)\Gk\left(\frac d2-\bk\right)\Gk\left(\ak+\bk-\frac d2\right)}{\Gk(\ak)\Gk(\bk)\Gk(d-\ak-\bk)}\,.
\end{equation}
twice, and then taken $\dk\rightarrow0$.  The result is finite, and so does not contribute to the anomalous dimension of $\boldsymbol\epsilon_i$. 

The final diagram for us to consider is:
\begin{equation}\label{PiEkEk}
\Pi_{\epsilon\epsilon}(x) = \feynmandiagram [inline=(a.base), horizontal = a to c] {
a [particle=$\epsilon(x)$] --  [scalar] b  -- [scalar] c [particle=$\epsilon(0)$],
b -- [half left] d -- [half left] b
}; = \frac{A_d(\Dk_\epsilon)}{2N}\int \frac{d^dz_1d^dz_2}{|z_1-z_2|^{2d-2\Dk_\epsilon}}\<\epsilon(x)\epsilon(0)\epsilon(z_1)\epsilon(z_2)\>\,.
\end{equation}
This diagram is almost identical to \eqref{PiSigSig} except that the external legs are now $\epsilon$ operators rather than $\sigma$ operators. We evaluate it in an identical fashion, finding that
\begin{equation}\begin{split}
\Pi_{\epsilon\epsilon}(x) &= \frac {A_d(\Dk_\epsilon)\zeta_\epsilon\log\left(|x|^2\right)}{N|x|^{2\Dk_\epsilon}}\,.
\end{split}\end{equation}
where
\begin{equation}
\zeta_\epsilon = S_d\int \frac{d^dw}{|w-\hat e|^{2d-2\Dk_\epsilon}}\,\<\epsilon(0)\epsilon(\infty)\epsilon(\hat e)\epsilon(w)\>_{\text{conn}}\,.
\end{equation}
Just like $\zeta_\sk$, we compute $\zk_\epsilon$ by expanding $\<\epsilon\epsilon\epsilon\epsilon\>$ as a sum of conformal blocks using \cite{Simmons-Duffin:2016wlq}, and then integrate the blocks numerically using \cite{Komargodski:2016auf}, finding that
\begin{equation}
\zeta_\epsilon \approx 2210.1\pm0.8\,.
\end{equation}
Combining everything together,
\begin{equation}
\<\boldsymbol\epsilon_1(x)\boldsymbol\epsilon_1(0)\>_{\text{conn}} = \frac1{|x|^{2\Dk_\epsilon}}\left(1-\frac1N +  \frac {A_d(\Dk_\epsilon)\zeta_\epsilon\log\left(|x|^2\right)}{2N} + O(N^{-2})\right)\,,
\end{equation}
and so we conclude that
\begin{equation}\label{bepsAnom}
\Dk^*_{\boldsymbol\epsilon}(N) = \Dk_\epsilon - \frac{A_d(\Dk_\epsilon)\zeta_\epsilon}{2N} + \dots \approx \Dk_\epsilon - \frac{0.2125}N + \dots\,.
\end{equation}

Now let us turn to $\fcy E(x)$, which at the cubic fixed point becomes the operator $\fk(x)$. We already wrote the leading corrections to $\<\fk(x)\fk(0)\>_{\text{conn}}$ in \eqref{2ptDiag}, and now must simply compute the two diagrams. We begin with the first diagram in \eqref{2ptDiag}:
\begin{equation}
\feynmandiagram [inline=(b.base), horizontal = a to d] {
a [particle = $\fk(x)$]  --  b  -- [half left]  c --  d [particle = $\fk(0)$],
c -- [half left] b
};  = \frac{f_{\epsilon\epsilon\epsilon}^2A_d(\Dk_\epsilon)^4}{2N^2} \int \frac{d^dz_1\dots d^dz_6}{\left(z_{24}z_{35}|x-z_1| |z_6|\right)^{2d-2\Dk_\epsilon}\left(z_{12}z_{13}z_{23}z_{45}z_{46}z_{56}\right)^{\Dk_\epsilon}}\,.
\end{equation} 
To evaluate the integrals, we make repeated use of the star-triangle relation
\begin{equation}\begin{split}\label{3ptInt}
&\int d^dz \frac 1 {(z-x_1)^{2a_1}(z-x_2)^{2a_2}(z-x_3)^{2a_3}} \\
&\qquad\qquad= \frac{\pi^{d/2}}{x_{12}^{d-2a_3}x_{23}^{d-2a_1}x_{13}^{d-2a_2}} \frac{\Gk(d/2-a_1)\Gk(d/2-a_2)\Gk(d/2-a_3)}{\Gk(a_1)\Gk(a_2)\Gk(a_3)}\,,
\end{split}\end{equation}
which holds when $a_1 + a_2 + a_3 = d$. We first integrate $z_1$ and $z_6$ using \eqref{3ptInt}, then next integrate $z_2$ and $z_5$, before finally integrating $z_3$. Once the dust settles, we find that
\begin{equation}\begin{split}
\feynmandiagram [inline=(b.base), horizontal = a to d] {
a [particle = $\fk(x)$] --  b  -- [half left]  c --  d [particle = $\fk(0)$],
c -- [half left ] b
};  &= \frac{f_{\epsilon\epsilon\epsilon}^2}{2\pi^{3d/2}N^2}\frac{\Gk\left(d-\frac{3\Dk_\epsilon}2\right)}{\Gk\left(\frac{3\Dk_\epsilon}2-\frac d2\right)}\left(\frac{\Gk(\Dk_\epsilon)\Gk(\frac{d-\Dk_\epsilon}2)}{\Gk(\frac{\Dk_\epsilon}2)\Gk(\Dk_\epsilon-\frac d2)}\right)^3 \\
&\times \frac 1 {|x|^{d-2\Dk_\epsilon}}\int \frac{d^dz_4}{|x-z_4|^d|z_4|^d}\,.
\end{split}\end{equation}
We can perform our final integral over $z_4$ using \eqref{2ptInt}, however, when we do so we find it diverges. To extract the finite contribution we use dimensional regularization:
\begin{equation}
\int \frac{d^{d+\dk} z_4}{|x-z_4|^{d}|z_4|^{d}} = \frac{\pi^{\frac{d+\dk}2}\Gk(\frac{d-\dk}2)\Gk(\frac\dk 2)^2}{\Gk(\dk)\Gk(\frac d2)^2|x|^{d-\dk}} \underset{\dk\rightarrow0}\longrightarrow \text{divergent} + \frac{4\pi^{d/2}\log |x|}{\Gk(\frac d2)|x-y|^d} \,.
\end{equation}
Subtracting the divergent term, we conclude
\begin{equation}
\feynmandiagram [inline=(b.base), horizontal = a to d] {
a [particle = $\fk(x)$] --  b  -- [half left]  c --  d [particle = $\fk(0)$] ,
c -- [half left ] b
}; = \frac{f_{\epsilon\epsilon\epsilon}^2B_d(\Dk_\epsilon)}{N^2}\frac{A_d(\Dk_\epsilon)\log|x|}{|x|^{2d-2\Dk_\epsilon}} \,,
\end{equation} 
where we define
\begin{equation}
B_d(\Dk) = -\frac{2\Gk(d-\frac{3\Dk}2)\Gk(\Dk-\frac d2)\Gk(\Dk)^2\Gk(\frac{d-\Dk}2)^3}{\Gk(\frac d2)\Gk(d-\Dk)\Gk(\frac{3\Dk-d}2)\Gk(\frac d2-\Dk)^2\Gk(\frac \Dk 2)^3}\,.
\end{equation}

Now we turn to the second diagram in \eqref{2ptDiag}:
\begin{equation}\begin{split}\label{diag2}
\feynmandiagram [inline=(b.base), horizontal = a to d] {
a [particle=$\fk(x)$]  --  b  -- [half left]  c -- [half left]   b -- d [particle=$\fk(0)$] 
}; &= \frac{A_d(\Dk)^3}{2N^2}\int d^dz_1\dots d^d z_4 \frac{\<\epsilon(z_1)\epsilon(z_2)\epsilon(z_3)\epsilon(z_4)\>_{\text{conn}}}{\left(z_{34}|x-z_1||z_2|\right)^{2d-2\Dk_\epsilon}} \,.
\end{split}\end{equation}
To evaluate this, we note that this diagram is just the shadow transform of \eqref{PiEkEk}, and so
\begin{equation}\begin{split}
\feynmandiagram [inline=(b.base), horizontal = a to d] {
a [particle=$\fk(x)$]  --  b  -- [half left]  c -- [half left]   b -- d [particle=$\fk(0)$] 
}; &= \frac{A_d(\Dk)^2}N \int d^dz_1d^dz_2 \frac{\Pi_{\epsilon\epsilon}(z_1-z_2)}{|x-z_1|^{2d-2\Dk_\epsilon}|z_2|^{2d-2\Dk_\epsilon}} \\
&= \frac{A_d(\Dk)^3\zeta_\epsilon}{N^2} \int d^dz_1d^dz_2 \frac{\log |z_1-z_2|}{|x-z_1|^{2d-2\Dk_\epsilon}|z_2|^{2d-2\Dk_\epsilon}|z_1-z_2|^{2\Dk_\epsilon}} \\
&= \frac{A_d(\Dk)^3\zeta_\epsilon}{N^2} \frac{\nb}{\nb\dk}\int d^dz_1d^dz_2 \frac{1}{|x-z_1|^{2d-2\Dk_\epsilon}|z_2|^{2d-2\Dk_\epsilon}|z_1-z_2|^{2\Dk_\epsilon-\dk}}\bigg|_{\dk=0}\,. 
\end{split}\end{equation}
We can now compute the integrals over $z_1$ and $z_2$ using \eqref{2ptInt}, and so find that
\begin{equation}\begin{split}
\feynmandiagram [inline=(b.base), horizontal = a to d] {
a [particle=$\fk(x)$]  --  b  -- [half left]  c -- [half left]   b -- d [particle=$\fk(0)$] 
}; = -\frac{A_d(\Dk)^2\zeta_\epsilon}{N^2} \frac{\log|x|}{|x|^{2d-2\Dk_\epsilon}}\,.
\end{split}\end{equation}

Having evaluated both diagrams in \eqref{2ptDiag}, we can combine them to find that:
\begin{equation}
\<\fk(x)\fk(0)\>_{\text{conn}} = \frac{A_d(\Dk_\epsilon)}{N|x|^{2d-2\Dk_\epsilon}}\left(1 + \frac{\log|x|}{N}\left(f_{\epsilon\epsilon\epsilon}^2B_d(\Dk_\epsilon) - A_d(\Dk_\epsilon) \zeta_\epsilon\right) + \dots\right)\,.
\end{equation}
We thus conclude that at large $N$, the dimension of $\fcy E(x) \propto \phi(x)$ is
\begin{equation}\begin{split}
\Dk_{\fcy E}^* &= d-\Dk_\epsilon + \frac1{2N}\left(  A_d(\Dk_\epsilon)\zeta_\epsilon - f_{\epsilon\epsilon\epsilon}^2B_d(\Dk_\epsilon)\right) + O(N^{-2}) \\
&\approx 1.5874 + \frac{0.0796}{N} + O(N^{-2})\,.
\end{split}\end{equation}

In Table~\ref{ETABLE}, we compare our large $N$ expression for $\Dk_{\fcy E}^*$ to calculations performed in both the $4-\epsilon$ expansion \cite{Kleinert95} and the $d=3$ fixed dimension expansion \cite{Varnashev00,Carmona99}. These expansions give series which are only asymptotically convergent and so require resummation.  As we can see from the table, for $N>3$ our results for $\Dk_{\fcy E}^*$ are in good agreement with those computed from the $d=3$ 4-loop and 6-loop expansions, but are a little smaller than estimates from the $4-\epsilon$ expansion at $5$-loop.  For $N=3$, the large $N$ result is considerably larger than any of the perturbative results. In this case we have an independent estimate of the conformal dimension of $\Dk_{\fcy E}^*$ from the $O(3)$ fixed point. Conformal dimensions in the $O(3)$ fixed point are expected to be very close to those of the $N=3$ cubic fixed point, with\footnote{Note that the critical exponent $\nu$ is defined by the equation $\nu_{\text{cubic}}^{-1} = d-\Dk_{\fcy E}^*$.} 
\begin{equation}
\nu_{O(3)}-\nu_{\text{cubic}} = 0.0003(3)
\end{equation}
estimated using the $d=3$ 6-loop expansion \cite{Calabrese02}. Combining this with the current most precise determination of this critical exponent, $\nu_{O(3)} = 0.71164(10)$, which was computed using Monte Carlo \cite{PhysRevB.102.024406}, we arrive at the result
\begin{equation}
\Dk^*_{\fcy E, N=3} = 1.5943(6)\,.
\end{equation}
This value is a little larger than the estimates from the perturbative expansions, and a little smaller than the estimates from the large $N$ expansion.

\begin{table}[t]
\begin{center}
{\renewcommand{\arraystretch}{1.2}
\begin{tabular} {|r| c c | c | c c c | c |}\hline       
          & \multicolumn{2}{c|}{$4-\epsilon$ at $5$-loop}    & $d=3$ at $4$-loop   &                     \multicolumn{3}{c|}{$d=3$ at $6$-loop}                 & Large $N$ \\ \hline
 $N$      & \cite{PhysRevE.58.5371} & \cite{Carmona:1999rm} & \cite{Varnashev00} & \cite{Carmona:1999rm}  & \cite{Varnashev00}  &  \cite{PhysRevB.62.12195}  & this work \\ \hline
 3        & 1.571(8)                & 1.573(8)               & 1.571               & 1.584(12)               & 1.580(8)             &  1.582(2)                  & 1.614     \\
 4        & 1.616(4)                & 1.617(4)               & 1.598               & 1.599(16)               & 1.601(10)            &  1.609(4)                  & 1.607     \\
 5        & 1.628(3)                & ---                    & 1.602               & ---                     & ---                  &  1.611(8)                  & 1.603     \\
 6        & 1.630(3)                & ---                    & 1.603               & ---                     & ---                  &  1.607(6)                  & 1.601     \\
 8        & ---                     & 1.617(4)               & 1.602               & 1.596(12)               & 1.600(7)             & 1.601(6)                   & 1.597     \\ 
 $\infty$ & 1.593(2)                & 1.594(4)               & ---                 & 1.588(16)               & 1.590(6)             & 1.588(2)                   & 1.587     \\  \hline
\end{tabular}}
\caption{Estimates for the conformal dimension of $\fcy E$ at the cubic fixed point, computing using the $4-\epsilon$ expansion at 5-loop, the fixed dimension expansion at 4 and 6-loop, and the large $N$ expansion.}
\label{ETABLE}
\end{center}
\end{table}

\subsubsection{Stress Tensor}

We now turn to the stress tensor. In the Ising model, or indeed, any local CFT, the stress tensor satisfies the Ward identity \cite{Osborn:1993cr}
\begin{equation}\label{wardId}
\left\<\nb_\mu T^{\mu\nu}(x)\fcy O_1(y_1)\dots \fcy O_n(y_n)\right\> = -\sum_{i = 1}^n \left\<\fcy O_1(y_1)\dots\nb^\nu\fcy O_i(y_i)\dots\fcy O_n(y_n)\right\>\,,
\end{equation}
while the two-point function is given by
\begin{equation}\begin{split}
\<T^{\mu\nu}(x)T^{\sk\rk}(0)\> &= \frac{(d-1)c_T}{2dS_d^2|x|^{2d}} \left(I^{\mu\rk}(x)I^{\nu\sk}(x) + (\mu\leftrightarrow\nu) - \frac2d \dk_{\mu\nu}\dk_{\rk\sk}\right)\,, \\
\text{ with } I^{\mu\nu}(x) &= \dk^{\mu\nu} - \frac{2x^\mu x^\nu}{x^2}\,.
\end{split}\end{equation}
With this choice of normalization, the quantity $c_T$, sometimes called the ``central charge'' of the theory, is defined such that $c_T = 1$ for a free massless scalar. Numeric bootstrap has been used to compute $c_T \approx 0.946534(11)$ for the 3d Ising model \cite{El-Showk:2014dwa}.

Each of the $N$ copies of the Ising models contributes an operator $T^{\mu\nu}_i(x)$, but once we couple the models together only the sum of these stress tensors
\begin{equation}
\fcy T^{\mu\nu}(x) = \sum_{i = 1}^N T^{\mu\nu}_i(x)
\end{equation}
remains conserved. The other $N-1$ linearly-independent operators
\begin{equation}
{\mathbf T}^{\mu\nu}_i(x) = T^{\mu\nu}_i(x) - \frac1N \fcy T^{\mu\nu}(x)  \qquad \text{ for } i = 1\,,\dots\,,N-1
\end{equation}
will acquire an anomalous dimension, so that $\nb{\mathbf T}^{\mu\nu}_i(x)\propto V^\nu$ eats some vector multiplet $V^\nu$. As we shall see, this multiplet recombination allows us to compute the anomalous dimension of ${\mathbf T}^{\mu\nu}_i$ without computing the $\<{\mathbf T}{\mathbf T}\>$ two-point function directly. Our calculation is similar to one performed in \cite{Behan:2259464}, in which the anomalous dimension of the stress tensor was computed in the presence of a long-ranged interaction.

We begin by using conformal invariance to fix
\begin{equation}\begin{split}
\<{\mathbf T}_i^{\mu\nu}(x){\mathbf T}_j^{\sk\rk}(0)\> = \delta_{ij}\frac{(d-1)c_T}{2dS_d^2|x|^{2\Dk_{\mathbf T}}} \left(I^{\mu\rk}(x)I^{\nu\sk}(x) + (\mu\leftrightarrow\nu) - \frac2d \dk_{\mu\nu}\dk_{\rk\sk}\right)\,,
\end{split}\end{equation}
and so find that
\begin{equation}\label{nbnbT}\begin{split}
\<\nb_\mu{\mathbf T}_i^{\mu\nu}(x)\nb_\nu{\mathbf T}_j^{\sk\rk}(0)\> = \delta_{ij}\frac{(\Dk_{\mathbf T}-d)(d-1)(d^2+d-2)c_T}{d^2S_d^2} \frac{I^{\nu\rk}(x)}{|x|^{2\Dk_{\mathbf T}+2}}\,.
\end{split}\end{equation}

But we know that because $\nb_\mu{\mathbf T}_i^{\mu\nu}(x)$ vanishes at infinite $N$, in order for it to be non-zero at finite $N$ it must eat some independent conformal primary $V^\nu(x)$ of dimension 4. The only such operator available is 
\begin{equation}
{\mathbf V}_i^\mu(x) = \fk(x)\nb_\nu{\boldsymbol\epsilon_i} - \frac{\Dk_\epsilon}{d-\Dk_\epsilon} {\boldsymbol\epsilon_i}\nb_\nu\fk(x)\,,
\end{equation}
and so we conclude that
\begin{equation}
\nb_\mu {\mathbf T}^{\mu\nu}_i(x) = b(N) {\mathbf V}_i^\nu(x)
\end{equation}
for some function $b(N)$. To fix $b(N)$, we first compute that
\begin{equation}\begin{split}\label{easyB}
\left\<\nb_\mu{\mathbf T}_i^{\mu\nu}(x)V^\lk(0)\right\> &= b(N)\left\<V^\nu(x)V^\lk(0)\right\> \\
&= \frac{2d(d-\Dk_\epsilon)A_d(\Dk_\epsilon)b(N)}{\Dk_\epsilon N} \frac{I^{\nu\lk}(x)}{|x|^{2d+2}} +O(N^{-2})\,.
\end{split}\end{equation}
But at large $N$ we can alternatively calculate $\left\<\nb_\mu{\mathbf T}_i^{\mu\nu}(x)V^\lk(0)\right\>$ diagrammatically. We begin by computing
\begin{equation}\label{wComp}\begin{split}
\left\<\nb_\mu{\mathbf T}_i^{\mu\nu}(x)\fk(y){\boldsymbol\epsilon}_i(0)\right\> &= \feynmandiagram [inline=(b.base), horizontal = a to b] {
a [particle=$\nb_\mu{\mathbf T}_i^{\mu\nu}(x)$]  -- [scalar]  b  -- [scalar]  c [particle=${\boldsymbol\epsilon}_i(0)$],
b -- d [particle=$\fk(y)$] 
}; + O(N^{-2}) \\
&= \frac{A_d(\Dk_\epsilon)}{N}\int d^dz \frac{\left\<\nb_\mu T^{\mu\nu}(x)\epsilon(z)\epsilon(0)\right\>}{|z-y|^{2d-2\Dk_\epsilon}} + O(N^{-2}) \\
&= -\frac{A_d(\Dk_\epsilon)}{N}\frac1{|x-y|^{2d-2\Dk_\epsilon}}\nb^\nu\frac 1{|x|^{2\Dk_\epsilon}} + O(N^{-2})\,,
\end{split}\end{equation}
where in passing from the second to the third line we have made use of the Ward identity \eqref{wardId}. We now use \eqref{wComp} to compute:
\begin{equation}\begin{split}\label{hardB}
\left\<\nb_\mu{\mathbf T}_i^{\mu\nu}(x)V^\lk(0)\right\> &= \left\<\nb_\mu{\mathbf T}_i^{\mu\nu}(x)\fk(0)\nb^\lk{\boldsymbol\epsilon}_i(0)\right\>- \frac{\Dk_\epsilon}{d-\Dk_\epsilon}\left\<\nb_\mu{\mathbf T}_i^{\mu\nu}(x){\boldsymbol\epsilon}_i(0)\nb^\lk\fk(0)\right\> \\
&= \frac{2(d-\Dk_\epsilon)A_d(\Dk_\epsilon)}{N} \frac{I^{\nu\lk}(x)}{|x|^{2d+2}} + O(N^{-2})\,.
\end{split}\end{equation}
Comparing this identity to \eqref{easyB}, we deduce that
\begin{equation}
b(N) = \frac{\Dk_\epsilon}{d}\,.
\end{equation}
We now use the multiplet recombination to compute
\begin{equation}\begin{split}
\<\nb_\mu{\mathbf T}_i^{\mu\nu}(x)\nb_\nu{\mathbf T}_j^{\sk\rk}(0)\> &= b(N)^2\left\<{\mathbf V}_i^\nu(x){\mathbf V}_j^\rk(0)\right\> \\
&= \frac{2(d-\Dk_\epsilon)\Dk_\epsilon A_d(\Dk_\epsilon)}{d^2 N} \frac{I^{\nu\rk}(x)}{|x|^{2\Dk_{\mathbf T}+2}} + O(N^{-2})\,,
\end{split}\end{equation}
and by comparing to the general result \eqref{nbnbT}, conclude that
\begin{equation}
\Dk_{\mathbf T} = d + \frac{2dS_d^2A_d(\Dk_\epsilon)\Dk_\epsilon(d-\Dk_\epsilon)}{c_T(d-1)^2(d+2)N} + \dots\,.
\end{equation}
Specializing to the case $d = 3$, we find that
\begin{equation}\begin{split}
\Dk_{\mathbf T} &= 3 - \frac{12\cot(\Dk_\epsilon \pi)}{5\pi c_T N}\Dk_\epsilon(\Dk_\epsilon-1)(\Dk_\epsilon-2)(\Dk_\epsilon-3)(2\Dk_\epsilon-3) + O(N^{-2}) \\
&\approx 3 + \frac{0.02159}{N} + \dots\,.
\end{split}\end{equation}

\subsection{Multi-trace Operators}
Having computed the leading corrections to the scaling dimensions of all single trace operators in the cubic fixed point with dimension $\Delta\leq3$, we now will move to consider multi-trace operators. 

Let us begin by considering the double-trace operators $\sk_i\sk_j$ for $i\neq j$, which together form an irreducible representation of the $S_N\ltimes \mathbb Z_2^N$ hypercubic symmetry. To compute their anomalous dimension, we expand
\begin{equation}\begin{split}\label{skSq}
\<\sk_1\sk_2(x)&\sk_1\sk_2(0)\>_{\text{conn}} - \<\sk_1(x)\sk_1(0)\>_{\text{conn}}\<\sk_2(x)\sk_2(0)\>_{\text{conn}} \\
&=  \feynmandiagram [inline=(b.base), horizontal = b to e] {
a [particle=$\sk_1(x)$] --  [scalar] b  -- [scalar] c [particle=$\sk_1(0)$],
b --  e,
f [particle=$\sk_2(x)$] --  [scalar] e  -- [scalar] g [particle=$\sk_2(0)$]
}; + O(N^{-2})\\
&= \frac{f_{\sk\sk\epsilon}^2A_d(\Dk_\epsilon)}{N|x|^{4\Dk_\sk-2\Dk_\epsilon}}\int\frac{d^dz_1 d^{d+\dk}z_2}{z_{12}^{2d-2\Dk_\epsilon}\left(|x-z_1||x-z_2||z_1||z_2|\right)^{\Dk_\epsilon}} + O(N^{-2})\,.
\end{split}\end{equation}
where we have introduced $\dk$ as a regulator. Using \eqref{3ptInt} to evaluate the $z_1$ integral and then \eqref{2ptInt} to evaluate the $z_2$ integral, we find that
\begin{equation}\begin{split}
\feynmandiagram [inline=(b.base), horizontal = b to e] {
a [particle=$\sk_1(x)$] --  [scalar] b  -- [scalar] c [particle=$\sk_1(0)$],
b --  e,
f [particle=$\sk_2(x)$] --  [scalar] e  -- [scalar] g [particle=$\sk_2(0)$]
}; &= \text{divergent} - \frac{f_{\sk\sk\epsilon}^2 C_d(\Dk_\epsilon)}{N}\frac{\log(|x|^2)}{|x|^{4\Dk_\sk}} \,,\\
\text{ where } C_d(\Dk) &= \frac{2\Gk\left(\frac{d-\Dk}2\right)^2\Gk(\Dk)}{\Gk\left(\frac d2\right)\Gk\left(\frac d2-\Dk\right)\Gk\left(\frac\Dk2\right)^2}\,.
\end{split}\end{equation}
Combining this with the expression for $\<\sk_1(x)\sk_1(0)\>_{\text{conn}}$ derived in Section \ref{sigAnom}, we find that
\begin{equation}
\<\sk_1\sk_2(x)\sk_1\sk_2(0)\>_{\text{conn}}  = \frac1{|x|^{4\Dk_\sk}}\left(1+\frac{\log(|x|^2)}{N}\left(A_d(\Dk_\epsilon)\zeta_\sigma - f_{\sigma\sigma\epsilon}^2C_d(\Dk_\epsilon)\right) + O(N^{-2})\right)\,,
\end{equation}
and so conclude that
\begin{equation}\begin{split}
\Dk_{\sk^2}^* &= 2\Dk_\sk + \frac1N\left(f_{\sigma\sigma\epsilon}^2C_d(\Dk_\epsilon)-A_d(\Dk_\epsilon)\zeta_\sigma \right) + \dots \\
&\approx 1.0363  + \frac{0.1684}N + \dots\,.
\end{split}\end{equation}
We should note that the splitting between single and double-trace operators under a double-trace deformation was previously computed in \cite{Giombi:2018vtc}, by first considering the leading $1/N$ corrections to four-point functions and then extracting double-trace anomalous dimension by decomposing these four-point functions into conformal blocks. Their results applied to the cubic fixed point imply that
\begin{equation}
\Dk_{\sk^2}^* - 2\Dk_{\sk}^* = \frac{f_{\sigma\sigma\epsilon}^2}N C_d(\Dk_\epsilon) + O (N^{-2})\,,
\end{equation}
which matches with our own calculation.

Now let us consider the more general tensor $\sk_{i_1}\dots\sk_{i_k}(x)$ for $k<N$, where each $i_m\neq i_n$ for any $m\neq n$. Such tensors again from an irreducible representation of $S_N\ltimes\mathbb Z_2^N$. It is not hard to see that at leading order in the $1/N$ expansion, the only diagrams contributing to $\<\sk_{i_1}\dots\sk_{i_k}(x)\sk_{i_1}\dots\sk_{i_k}(0)\>_{\text{conn}}$ are products of \eqref{PiSigSig} or \eqref{skSq} with disconnected $\<\sk\sk\>$ two-point functions, so that
\begin{equation}\begin{split}
\big\<\sk_{i_1}\dots\sk_{i_k}(x)&\sk_{i_1}\dots\sk_{i_k}(0)\big\>_{\text{conn}} = \frac{k}{|x|^{2(k-1)\Dk_\sk}} \feynmandiagram [inline=(b.base), horizontal = a to c] {
a [particle=$\sk(x)$] --  [scalar] b  -- [scalar] c [particle=$\sk(0)$],
b -- [half left] d -- [half left] b
}; \\
&+ \frac{k(k-1)}{2|x|^{2(k-2)\Dk_\sk}} \feynmandiagram [inline=(b.base), horizontal = b to e] {
a [particle=$\sk_1(x)$] --  [scalar] b  -- [scalar] c [particle=$\sk_1(0)$],
b --  e,
f [particle=$\sk_2(x)$] --  [scalar] e  -- [scalar] g [particle=$\sk_2(0)$]
}; + O(N^{-2})\,.
\end{split}\end{equation}
We therefore conclude that
\begin{equation}\begin{split}
\Dk_{\sk^k}^* &= k\Dk_\sk + \frac1N\left(\frac{k(k-1)}2f_{\sk\sk\epsilon}^2C_d(\Dk_\epsilon) - \frac k2 \zeta_\sigma A_d(\Dk_\epsilon) \right) + \dots \\
&\approx 0.5181k + \frac {0.0836 k^2-0.0830 k}N + \dots\,.
\end{split}\end{equation}

Next we turn to the tensor ${\boldsymbol\epsilon}_i{\boldsymbol\epsilon}_j(x)$, calculating 
\begin{equation}\begin{split}\label{ekSq}
\<\boldsymbol\epsilon_1\boldsymbol\epsilon_2(x)&\boldsymbol\epsilon_1\boldsymbol\epsilon_2(0)\>_{\text{conn}} - \<\boldsymbol\epsilon_1(x)\boldsymbol\epsilon_1(0)\>_{\text{conn}}\<\boldsymbol\epsilon_2(x)\boldsymbol\epsilon_2(0)\>_{\text{conn}} \\
&=  \feynmandiagram [inline=(b.base), horizontal = b to e] {
a [particle=$\epsilon_1(x)$] --  [scalar] b  -- [scalar] c [particle=$\epsilon_1(0)$],
b --  e,
f [particle=$\epsilon_2(x)$] --  [scalar] e  -- [scalar] g [particle=$\epsilon_2(0)$]
}; + O(N^{-2})\\
&= \frac{f_{\epsilon\epsilon\epsilon}^2A_d(\Dk_\epsilon)}{N|x|^{4\Dk_\epsilon-2\Dk_\epsilon}}\int\frac{d^dz_1 d^{d+\dk}z_2}{z_{12}^{2d-2\Dk_\epsilon}\left(|x-z_1||x-z_2||z_1||z_2|\right)^{\Dk_\epsilon}} + O(N^{-2})\,.
\end{split}\end{equation}
This leading correction is almost identical to the first diagram in \eqref{skSq}, differing only in the OPE coefficient and power of $|x|$ out front:
\begin{equation}\begin{split}
\feynmandiagram [inline=(b.base), horizontal = b to e] {
a [particle=$\epsilon_1(x)$] --  [scalar] b  -- [scalar] c [particle=$\epsilon_1(0)$],
b --  e,
f [particle=$\epsilon_2(x)$] --  [scalar] e  -- [scalar] g [particle=$\epsilon_2(0)$]
}; &=  \frac{f_{\epsilon\epsilon\epsilon}^2A_d(\Dk_\epsilon)}{N|x|^{2\Dk_\epsilon}}\int\frac{d^dz_1 d^{d+\dk}z_2}{z_{12}^{2d-2\Dk_\epsilon}\left(|x-z_1||x-z_2||z_1||z_2|\right)^{\Dk_\epsilon}}\\
&= \text{divergent} - \frac{f_{\epsilon\epsilon\epsilon}^2 C_d(\Dk_\epsilon)}{N}\frac{\log(|x|^2)}{|x|^{4\Dk_\epsilon}}
\end{split}\end{equation}
We thus conclude that 
\begin{equation}\begin{split}
\Dk_{\boldsymbol\epsilon^2}^* &= 2\Dk_\epsilon + \frac1N\left(f_{\epsilon\epsilon\epsilon}^2C_d(\Dk_\epsilon)-A_d(\Dk_\epsilon)\zeta_\epsilon \right) + \dots \\
&\approx 2.8253  - \frac{0.0702}N + \dots\,.
\end{split}\end{equation}
Once again, we can easily extend the calculation to the more general tensor $\boldsymbol\epsilon_{i_1}\dots\boldsymbol\epsilon_{i_n}$, finding that
\begin{equation}\begin{split}
\Dk_{\boldsymbol\epsilon^k}^* &= k\Dk_\epsilon + \frac1N\left(\frac{k(k-1)}2f_{\epsilon\epsilon\epsilon}^2C_d(\Dk_\epsilon)-\frac k2A_d(\Dk_\epsilon)\zeta_\epsilon \right) + \dots \\
&\approx 1.4125k + \frac{0.1775k^2-0.3900k}N + \dots\,,
\end{split}\end{equation}

\section{Discussion}
\label{DISC}

In this work we studied the large $N$ limit of the cubic fixed point. We derived a diagrammatic expansion for the $1/N$ corrections to correlators at the fixed point. Combining this with results from the 3d Ising model numeric bootstrap, we were able to compute leading corrections to the anomalous dimensions of a number of operators. 

Our results for $\Dk_{\fcy E}^*$ could be compared to perturbative calculations and found to be in good agreement when $N\geq4$. Unfortunately, the other conformal dimensions considered in this paper have to our knowledge never been accurately computed before,\footnote{As noted in Section~\ref{SIGANOM}, $\Delta_\sk$ has been studied in both the $4-\epsilon$ and $d=3$ perturbative expansions, but the uncertainties are large relative to the size of the $1/N$ correction to $\Dk_\sk^*$ derived in this paper. A number of other conformal dimensions have been computed at to 2-loop in the $4-\epsilon$ expansion \cite{Dey:2016mcs}, but this does not suffice to reliably extrapolate to $\epsilon=1$.} whether that be through perturbation theory, Monte Carlo simulation, or numeric conformal bootstrap. Recent numeric conformal bootstrap studies have considered theories with hypercubic symmetry \cite{Rong_2018,Stergiou18,Kousvos:2018rhl,Kousvos:2019hgc}, but these have so far been unable to isolate the cubic fixed point. It would be very interesting to try to use the large $N$ results derived here to determine reasonable assumptions that could be fed into the conformal bootstrap, hopefully allowing the cubic fixed point to be bootstrapped and hence our large $N$ results to be tested.

While we focused in this paper on the cubic fixed points, our methods are general and can apply to any theory with a vector-like large $N$ limit. The most straightforward generalization is to the $MN$ models \cite{PhysRevB.10.892}, which have $S^N\ltimes O(M)^N$ symmetry. These should be related to the $O(M)$ models via a renormalization group flow triggered by the interaction
\begin{equation}\label{MNCouple}
L_{\text{int}} = \int d^dx\, \lk\sum_{i\neq j}S_i(x)S_j(x)
\end{equation}
coupling $N$ $O(M)$ models, where $S(x)$ is the lowest dimension singlet operator in the $O(N)$ model. Unlike the Ising model, for $M\geq 2$ the dimension of $\Dk_S>\frac32$ when $d=3$, so that the interaction \eqref{MNCouple} is irrelevant. As a result, the $MN$ fixed points are not stable for any $M>1$. Nevertheless, we can still study the $MN$ model at large $N$, and, with sufficiently good numerical bootstrap results, could use the methods of this paper to extract leading corrections to anomalous dimensions at these fixed points. 

It is straightforward to generalize our results to theories with double-trace deformations of the form
\begin{equation}
L_{\text{int}} = \int d^dx\, \lk\sum_{i\neq j}\dk_{IJ}\fcy O_i^I(x)\fcy O^J_j(x)\,,
\end{equation}
where $\fcy O_I(x)$ are operators in a conformal field theory which are charged under some global symmetry $\fcy G$. For instance, the 3d $O(M)$ models for $M\geq 2$ contain a symmetric tensor $t_{ij}(x)$ whose conformal dimension $\Dk_t<\frac 32$. We can use this operator to construct a number of relevant double-trace deformations which break the $S_N\ltimes O(M)^N$ symmetry of $N$ decoupled $O(M)$ models down to discrete subgroups, and so can construct a large class of conformal field theories with tractable large $N$ limits.

\subsubsection*{Acknowledgments}

I would like to thank Connor Behan and Shai Chester for assistance with numeric computations, Andreas Stergiou for useful discussions, and Silviu Pufu for both useful discussions and for reading through the manuscript. I am supported in part by the Simons Foundation Grant No.~488653, and by the US NSF under Grant No.~1820651\@.  I am also supported in part by the General Sir John Monash Foundation.  I am grateful to the ANU Research School of Physics for their hospitality during the completion of this project.

\appendix

\section{Regulating the Hubbard--Stratonovich Trick}
\label{REGHS}

In this appendix we will discuss proper meaning of \eqref{initGen}. Naively, we would expect that
\begin{equation}\begin{split}
\int D\phi(x) &\exp\left(\int d^dx\,\frac1{2\lk}\fk(x)^2 + \fk(x)\sum_{i=1}^N\epsilon_i(x)\right) \\
&\propto \exp\left(-\frac{\lk}{2}\int d^dx\ d^dy\ \dk(x-y)\left(\sum_i\epsilon_i(x)\right)\left(\sum_i\epsilon_i(y)\right)\right) \\
&= \exp\left(-\frac{\lk}{2}\int d^dx\ \sum_{i,j}\epsilon_i(x)\epsilon_j(x)\right)\,,
\end{split}\end{equation}
however, this last expression is problematic as $\epsilon_i(x)^2$ is not a well-defined operators. To circumvent this, we modify the path-integral of $\fk(x)$ so that
\begin{equation}\<\fk(x)\fk(y)\> = R_\ek(x-y) \underset{\ek\rightarrow 0}\longrightarrow \dk(x-y)\,.\end{equation}
We then write
\begin{equation}\begin{split}
\int D\phi(x) &\exp\left(\int d^dx\,\frac1{2\lk}\fk(x)^2 + \fk(x)\sum_{i=1}^N\epsilon_i(x)\right) \\
&= \lim_{\epsilon\rightarrow0}\exp\left(-\frac{\lk}{2}\int d^dx\ d^dy\ R_\epsilon(x-y)\left(\sum_i\epsilon_i(x)\right)\left(\sum_i\epsilon_i(y)\right)\right)\,.
\end{split}\end{equation}
Using the OPE expansion, we can then evaluate
\begin{equation}\begin{split}
\int d^dy\, R_\epsilon(x-y)\epsilon_i(x)\epsilon_i(y) =   \sum_A f_{\epsilon\epsilon A}A(x)\int d^dz \frac{R_\epsilon(z)}{|z|^{2\Dk_\epsilon-\Dk_A}} + \text{descendants} 
\end{split}\end{equation}
We can ignore descendants, as these becomes total derivatives upon integration over $x$. In the limit $\epsilon\rightarrow0$, terms for which $\Dk_A>2\Dk_\epsilon$ will go to zero, whereas if $\Dk_A<2\Dk_\epsilon$, the coefficient diverge and need to be subtracted off. This means that we only have to worry about operators satisfying $\Dk_A = 2\Dk_\epsilon$, but of course no such operator exists in the 3d Ising model. We thus conclude that \eqref{initGen} is true so long as we regulate and subtract off divergences.

\printbibliography

\end{document}